\documentclass[final,5p,times,twocolumn]{elsarticle}

\usepackage{amssymb}
\usepackage{amsmath}
\usepackage{grffile}

\journal{Journal of Magnetism and Magnetic Materials}

\newcommand{\kk}{\mathbf{k}}
\newcommand{\I}{\mathrm{i}}

\begin{document}

\begin{frontmatter}



\title{Spiral magnetic order, non-uniform states and electron correlations in the conducting transition metal systems}


\author[Ekb,Urfu]{P.A. Igoshev}
\author[Izh]{M.A. Timirgazin}
\author[Izh]{A.K. Arzhnikov}
\author[Urfu]{T.V. Antipin}
\author[Ekb]{V.Yu. Irkhin}

\address[Ekb]{Institute of Metal Physics, Russian Academy of Sciences, 620990 Ekaterinburg, Russia}
\address[Izh]{Physical-Technical Institute, 426000, Kirov str. 132, Izhevsk, Russia}
\address[Urfu] {Ural Federal University, 620002 Ekaterinburg, Russia}

\ead{valentin.irkhin@imp.uran.ru}

\begin{abstract}
The ground-state magnetic phase diagram is calculated within the Hubbard  and   $s$-$d$ exchange (Kondo) 
for square and simple cubic lattices vs band filling and interaction parameter. 
The difference of the results owing to the presence of  localized moments in the latter model is discussed. 
We employ a generalized Hartree-Fock approximation (HFA) to treat commensurate ferromagnetic (FM), antiferromagnetic (AFM), and incommensurate (spiral) magnetic phases. 
The electron correlations are taken into account within the Hubbard model by using the Kotliar-Ruckenstein slave boson approximation~(SBA). 
The main advantage of this approach is a correct qualitative description of the paramagnetic phase: its energy becomes considerably lower as compared with HFA, and the gain in the energy of magnetic phases is substantially reduced. 
\end{abstract}


\end{frontmatter}


\section{Introduction}\label{sec:intro}

Magnetic properties of strongly correlated transition-metal systems and their relation to doping, lattice geometry and band structure are still extensively investigated. 
The general result of these investigations is the  existence of first-order transition between various commensurate and incommensurate magnetic states which invokes a phase separation  (first discussed by  Visscher~\cite{Visscher73}).

To describe the properties of such systems one uses many-electron models like the Hubbard, $s$-$d$ exchange (Kondo) model and Anderson lattice model. These  are widely applied, {\it e.~g.}, for high-$T_c$ cuprates and rare earth compounds. 
There exist some relations between these models in various parameter regions. 
The problem of local moments formation, {\it e.~g.} within the Hubbard model, is very difficult and still under investigation~\cite{Moriya}. 
On the other hand, in the  $s$-$d$ exchange model the localized moments (spins $S$) are explicitly present in the Hamiltonian (although they are screened in the Kondo regime). 

In the present paper we perform an investigation of the magnetic phase diagram of the Anderson-Kondo lattice model for the square and simple cubic lattices including the phase separation, as well as non-collinear magnetic ordering, and trace these relations. 
We treat the influence of inter-orbital interaction on the spiral state formation, the difference of the Hubbard (one-orbital) and Anderson-Kondo lattice (two-orbital) model results being considered.

\section{Theory}\label{sec:theory}

The theoretical investigation of spiral formation in itinerant systems is generally based (in minimal variant) on the non-degenerate Hubbard and the Anderson models. 
Within the Hubbard model 
\begin{equation}
      \label{eq:original_H}
      \mathcal{H}_{\rm H} = \sum_{ij\sigma} t_{ij}d^\dag_{i\sigma}d^{}_{j\sigma}+U\sum_i n^d_{i\uparrow}n^d_{i\downarrow},
\end{equation}
the itinerant electrons demonstrate both transport and interaction induced magnetic properties. 
Here the matrix elements of the electron transfer are $t_{ij} = -t$ for the nearest neighbors 
(we assume $t>0$), $d^\dag_{i\sigma},d^{}_{i\sigma}$ are the electron creation and annihilation
operators, respectively, $i$ is the site number, $\sigma=\uparrow,\downarrow$ is the spin projection, the last term being responsible for the one-site Coulomb interaction of $d$-electrons, $n^d_{i\sigma}=d^{\dagger}_{i\sigma}d_{i\sigma}$.

In the case of the Anderson model transport and magnetic properties are separated between different systems, $s$ and $d$ correspondingly: 
\begin{multline}\label{eq:hamiltonian of Anderson} 
\mathcal{H}_{\rm A} = \sum_{ij\sigma} t_{ij}c^{\dagger}_{i\sigma}c_{j\sigma} + \epsilon_d\sum_{i\sigma}d^{\dagger}_{i\sigma}d^{}_{i\sigma} \\
 + V\sum_{i\sigma}(c^{\dagger}_{i\sigma}d^{}_{i\sigma} + d^{\dagger}_{i\sigma}c^{}_{i\sigma}) +
U\sum_i n_{i\uparrow}^d n_{i\downarrow}^d,
\end{multline}
$c^{\dagger}_{i\sigma},c^{}_{i\sigma}$ is creation/annihilation electron operator in itinerant (`$s$-electron') state at site $i$. 
$\epsilon_d$ is the energy of localized (`$d$-electron') electron state, 
$V$ is one-site $s$-$d$ hybridization providing the coupling between these subsystems. 
The total electron concentration in the system is $n = n_s + n_d$, where $n_s = \sum_\sigma\langle c^{\dagger}_{i\sigma}c^{}_{i\sigma}\rangle$ and $n_d=\sum_\sigma\langle d^{\dagger}_{i\sigma}d^{}_{i\sigma}\rangle$ are the occupation numbers for itinerant and localized states respectively. 

Provided that the $d$-level is well below the Fermi energy and Coulomb interaction is sufficiently large ($|V|\ll \epsilon_{\rm F}-\epsilon_d$, $U$), this model can be reduced by the Schrieffer-Wolf transformation~\cite{SW}  to the $s$-$d$ exchange model 
with spin $S=1/2$ and the exchange parameter
\begin{equation}
	I = V^2[1/(\epsilon_d-\epsilon_{\rm F}) - 1/(U+\epsilon_d-\epsilon_{\rm F})],
\end{equation}
where $\epsilon_{\rm F}$ is the Fermi level. 
The Hamiltonian of the latter model reads
\begin{equation}\label{eq:sd_model_def}
	\mathcal{H}_{s-d} = \sum_{\kk\sigma} t_{\kk}c^\dag_{\kk\sigma}c^{}_{\kk\sigma} - I\sum_{i \sigma\sigma'} (\mathbf{S}_i\cdot\vec\sigma_{\sigma\sigma'}) c^\dag_{i\sigma}c^{}_{i\sigma'},
\end{equation}
$\mathbf{S}_i$ is localized spin operator, $\vec{\sigma}_{\sigma\sigma'}$ stands for Pauli matrices.

We consider  ferromagnetic and antiferromagnetic, as well as spiral incommensurate magnetic order,  with the magnetization 
$\mathbf{m}^{s,d}_i = \sum_{\sigma\sigma'}\langle(c,d)^\dag_{i\sigma}\vec{\sigma}_{\sigma\sigma'}(c,d)_{i\sigma'}\rangle$ being  modulated in the $xy$-plane with the wave vector $\mathbf{Q}$ \cite{Igoshev10}.
After local rotation in spin space matching the average magnetization direction at different sites we have a hopping matrix, non-diagonal with respect to spin, $t_{ij}\delta_{\sigma\sigma'}\rightarrow t_{ij}^{\sigma\sigma'}$~\cite{Fresard92}. 

The saddle-point expression for the spiral state grand canonical potential (per site) $\Omega$ has the form
\begin{equation}
	\label{eq:Omega_sdl_pt}
	\Omega = 
	\Omega_{\rm f} + \Omega_{\rm bg},
\end{equation}
where $\Omega_{\rm f}$ is a contribution from effective fermion Hamiltonian $\mathcal{H}_{\rm f}$ describing their motion in the ground state in some effective field,
\begin{equation}
	\label{eq:Omega_f_definition}
	\Omega_{\rm f} \equiv (1/N)\sum_{\kk \nu}(E_{\nu}(\kk)-\mu)f(E_{\nu}(\kk))
\end{equation}
where $E_{\nu}(\kk)$ are eigenvalues of $\mathcal{H}_{\rm f}$, $f(E) = \theta(\mu - E)$ is the Fermi function at $T = 0$, $\mu$ is the chemical potential, $N$ is the lattice site number.
$\Omega_{\rm bg}$ is `inner' effective field (`background') contribution to the grand canonical potential. 

Within the Hubbard model
\begin{equation}\label{eq:Hf_def}
	\mathcal{H}_{\rm f} = \sum_{\sigma\sigma'\kk} (z_\sigma z_{\sigma'}(e_+(\kk)\delta_{\sigma\sigma'}+e_-(\kk)\delta_{\sigma,-\sigma'})
	+\delta_{\sigma\sigma'}\lambda_\sigma) d^\dag_{\kk\sigma}d^{}_{\kk\sigma'},
\end{equation}
where
\begin{equation}
    \label{eq:esa}
    e_\pm(\kk) =(1/2)(t_{\kk+\mathbf{Q}/2}\pm t_{\kk-\mathbf{Q}/2}),
\end{equation}
\begin{equation}
	t_\kk = \sum_{i}\exp(\I\kk(\mathbf{R}_i-\mathbf{R}_j))t_{ij}
\end{equation}
is the bare electron spectrum. 
The concrete expressions for the spectrum renormalization factors $z^2_\sigma$, $\lambda_\sigma$ and $\Omega_{\rm bg}$ depend on the approximation employed.

The resulting wave vector $\bf{Q}$ is determined by  minimization of $\Omega$ over various spiral states 
at fixed $\mu$ 
which allows to take into account 
the phase separation possibility
~\cite{Igoshev10,Igoshev2013}.

\subsection{Hartree-Fock approximation}\label{sec:HF}
The generalized HFA for the Coulomb interaction in $d$-subsystem reads
\begin{equation}
	Un_{d\uparrow}n_{d\downarrow}\rightarrow Un^d_{i\uparrow}\langle n_{d\downarrow}\rangle + U\langle n^d_{i\uparrow}\rangle n^d_{i\downarrow} - U\langle n^d_{i\uparrow}\rangle\langle n^d_{i\downarrow}\rangle. 
\end{equation}
The main shortcoming of HFA is the account of contributions of singly and doubly occupied states to $n^d_{\sigma}\equiv\langle n^d_{i,\sigma}\rangle$ in the equal way, which becomes incorrect at sufficiently large $U$.  

Correlation-induced band narrowing is absent, $z^2_\sigma = 1$ and $\lambda_\sigma = Un^d_{-\sigma}$ in Eq.~(\ref{eq:Hf_def}), 
\begin{equation}\label{eq:Omega_bg}
	\Omega_{\rm bg} 
	= -Un^d_\uparrow n^d_\downarrow.
\end{equation}
In the case of the Anderson model we have
\begin{multline}\label{eq:Hf_A_def}
	\mathcal{H}_{\rm f} = \sum_{\sigma\sigma'\kk} (e_+(\kk)\delta_{\sigma\sigma'}+e_-(\kk)\delta_{\sigma,-\sigma'})
	c^\dag_{\kk\sigma}c^{}_{\kk\sigma'} \\
	+ V\sum_{\kk\sigma}(c^\dag_{\kk\sigma}d^{}_{\kk\sigma} + d^\dag_{\kk\sigma}c^{}_{\kk\sigma}) 
	+ \sum_{\kk\sigma}(\epsilon_d + U n_{d,-\sigma}) d^\dag_{\kk\sigma}d^{}_{\kk\sigma},
\end{multline}
so that two types of mixing are present: the hybridization $V$ of $s$- and $d$-systems and spin flip terms proportional to $e_-(\kk)$. 

\subsection{The account of correlations: slave boson approximation}\label{sec:SB}
Besides HFA, we apply SBA~\cite{Kotliar86} to the single-band Hubbard model. 
The idea of this approximation is extension of configuration space. This duplicates  the standard description based on the Slater determinant wave functions (related to operators $c$, $c^\dag$) by using the  boson operators $e_i$, $p_{i\sigma}$, $d_i$ and their conjugates which correspond to empty,  singly occupied, and  doubly occupied states respectively. 
The bosonic space construction is realized by  requiring the presence of exactly one boson at any time, 
\begin{equation}
	\label{eq:norm}
	e^\dag_i e^{}_i + p_{i\uparrow}^\dag p^{}_{i\uparrow} + p_{i\downarrow}^\dag p^{}_{i\downarrow}+d^\dag_i d^{}_i = 1.
\end{equation} 
Any on-site transition operator (Hubbard $X$ operator~\cite{Hubbard2}) has its counterpart in the slave boson language, {\it e.g.},~$X^{0\sigma}_i\sim \mathcal{P}e^\dag_i p_{i\sigma}\mathcal{P}$ for any site $i$, $\mathcal{P}$ being the projection operator onto the corresponding subspace. 
The exact coherence of fermion and boson systems is established by the replacement
\begin{equation}
	c^\dag_{i\sigma}c_{j\sigma'} \rightarrow g^{(1)}_{i\sigma}(p^\dag_{i\sigma}e_j + d_i^\dag p_{j,-\sigma'})g^{(2)}_{j\sigma'} c^\dag_{i\sigma}c_{j\sigma'},
\end{equation}
$g^{(1)}_{i\sigma}$~($g^{(2)}_{i\sigma}$) being an operator equal to unity on the subspace defined by the equation $d^\dag_i d^{}_i + p^\dag_{i,\sigma}p^{}_{i,\sigma} = 1$ ($e^\dag_i e^{}_i + p^\dag_{i,-\sigma}p^{}_{i,-\sigma} = 1$) needed to reproduce HFA results at small $U$ within the saddle-point approximation 
~\cite{Kotliar86}. 
The consequence of this coherence is the connection of site occupation numbers in terms of fermions and bosons,
\begin{equation}
	c^\dag_{i\sigma}c^{}_{i\sigma} = d^\dag_i d^{}_i + p^\dag_{i\sigma} p^{}_{i\sigma}. 
\end{equation}
The Hubbard on-site interaction becomes diagonal in the boson representation: $U n_{i\uparrow}n_{i\downarrow} = UX^{22}_i = U d^\dag_i d^{}_i$. 
Within the saddle-point approximation, the bosonic operators are replaced by $c$-numbers. This yields an improvement of the Hartree-Fock approximation,  so that the corresponding effective field can be interpreted as a result of average many-electron site  amplitudes satisfying  Eq. (\ref{eq:norm}). 
The partial electronic concentrations are parametrized by
\begin{equation}
	\label{eq:nd}
	n^d_\sigma = p^2_\sigma + d^2. 
\end{equation} 
The subband narrowing is the $c$-number function of extremal bosonic fields, 
\begin{equation}
	\label{eq:z_def2}
	z^2_\sigma=(1-d^2-p_\sigma^2)^{-1} (ep_\sigma+p_{\bar\sigma}d)^2 (1-e^2-p_{\bar\sigma}^2)^{-1}.
\end{equation}
The grand canonical background  potential has the form
\begin{equation}\label{eq:Omega_bg_sb}
	\Omega_{\rm bg} 
	= Ud^2 - \sum_\sigma\lambda_\sigma(p_\sigma^2+d^2).
\end{equation}
The impact of the averaged site states on electron states manifests itself in two types of renormalizations of bare spectrum: (i) narrowing of the bare spectrum,  similar to the Hubbard--I approximation\cite{Hubbard}, which is
specified by the factor $z^2_\sigma\le 1$, (ii) the additional energy shift $\lambda_\sigma$ which is an analogue of the Harris--Lange shift~\cite{Harris-Lange1967}.
Both these quantities are essentially spin dependent, which allows one to study the magnetic states formation.
Unlike  HFA, in the slave-boson approach  $\lambda_\sigma$ cannot be
expressed in terms of $n$ and $m$ only and 
is obtained from the saddle-point equation as 
\begin{equation}
	\label{eq:lmb_eq}	
	\lambda_\sigma = (ed-p_\uparrow p_\downarrow)\left[ \Phi_{\sigma}\left(\frac{p_{\bar\sigma}/e}{e^2+p_{\bar\sigma}^2} + \frac{d/p_{\sigma}}{p_{\sigma}^2+d^2}\right)+\Phi_{\bar\sigma}/(ep_\sigma)\right],
\end{equation}
\begin{equation}
	\label{eq:Phi_def}
	 \Phi_{\sigma}\equiv\frac{ep_\sigma+p_{\bar\sigma}d}{(e^2+p_{\bar\sigma}^2)(p_{\sigma}^2+d^2)}\frac1{N}\sum_{\kk\nu} f(E_\nu(\kk))\frac{\partial E_\nu(\kk)}{\partial (z^2_{\sigma})}. 
\end{equation}	
The balance of bosonic $p_\sigma$ and $d$ amplitudes is determined by the equation
\begin{equation}
	\label{eq:main_p}
	(ed-p_\uparrow p_\downarrow)\sum_\sigma \left(\frac1{ep_{\bar\sigma}}+\frac1{p_\sigma d}\right)\Phi_\sigma = U.
\end{equation}

The problem of correlations in the Anderson model can be also treated in SBA~\cite{Moller}. 
However, this yields the exponent in the Kondo temperature which is different from the value in the single impurity case. 
On the other hand, pseudofermion representation by Coleman and Andrei in the $s$-$d$ model~\cite{711} does not have such a drawback and can be successfully applied to the problem of magnetic ordering~\cite{608,Lacroix}.

\section{Results}\label{sec:results}

Here we consider the ground state magnetic phase diagram for square and simple cubic lattices for both the Hubbard and $s$-$d$ models calculated in the generalized HFA. For the Hubbard model we present also SBA results (in the case of bcc and fcc lattices, they can be found in Ref.~\cite{JPCM}).

The more general case of intermediate valence within the Anderson lattice model will be considered elsewhere, see also~\cite{Antipin}. 

\begin{figure}[h!]
\includegraphics[width=0.5\textwidth]{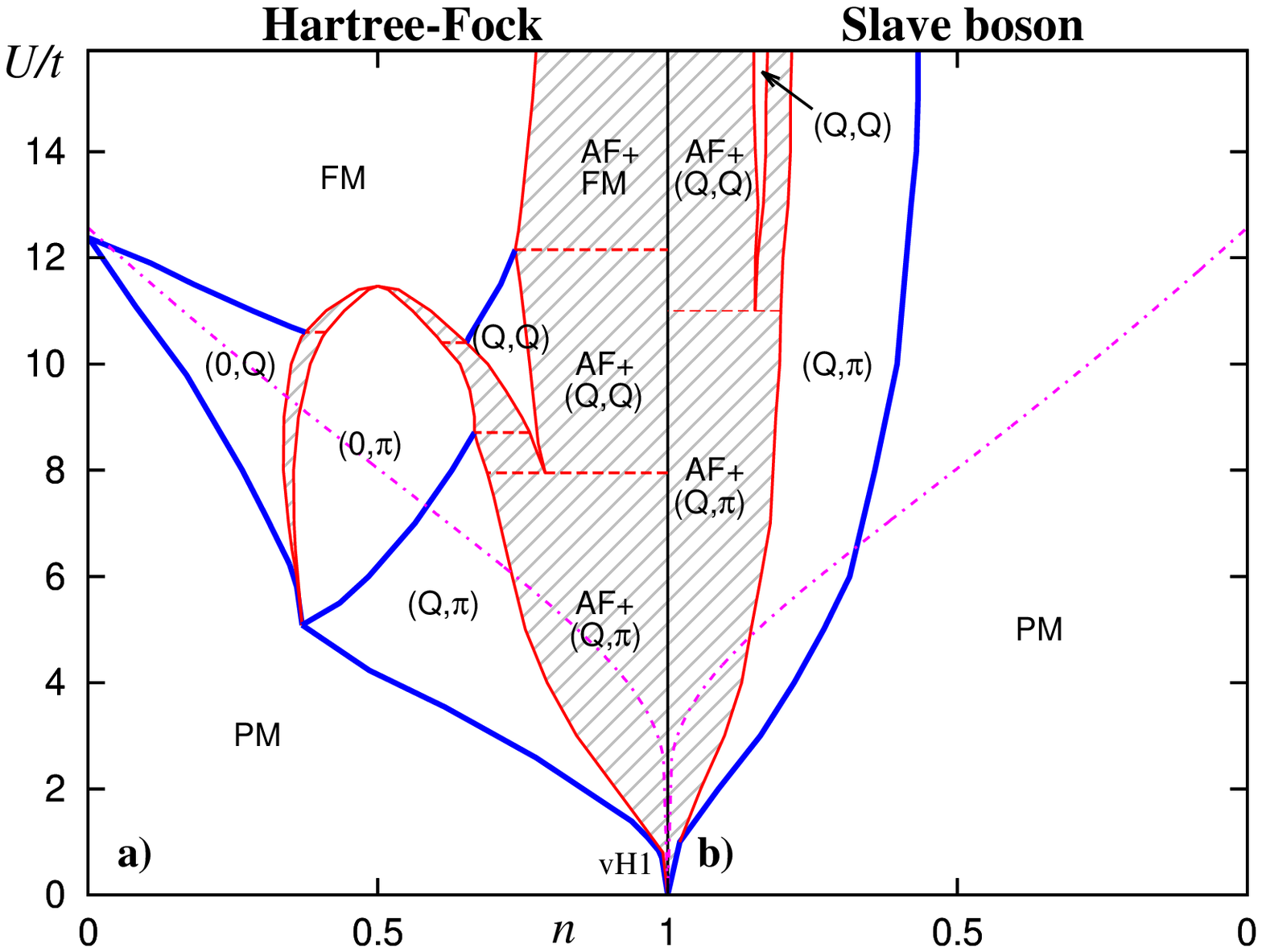}
\includegraphics[width=0.49\textwidth]{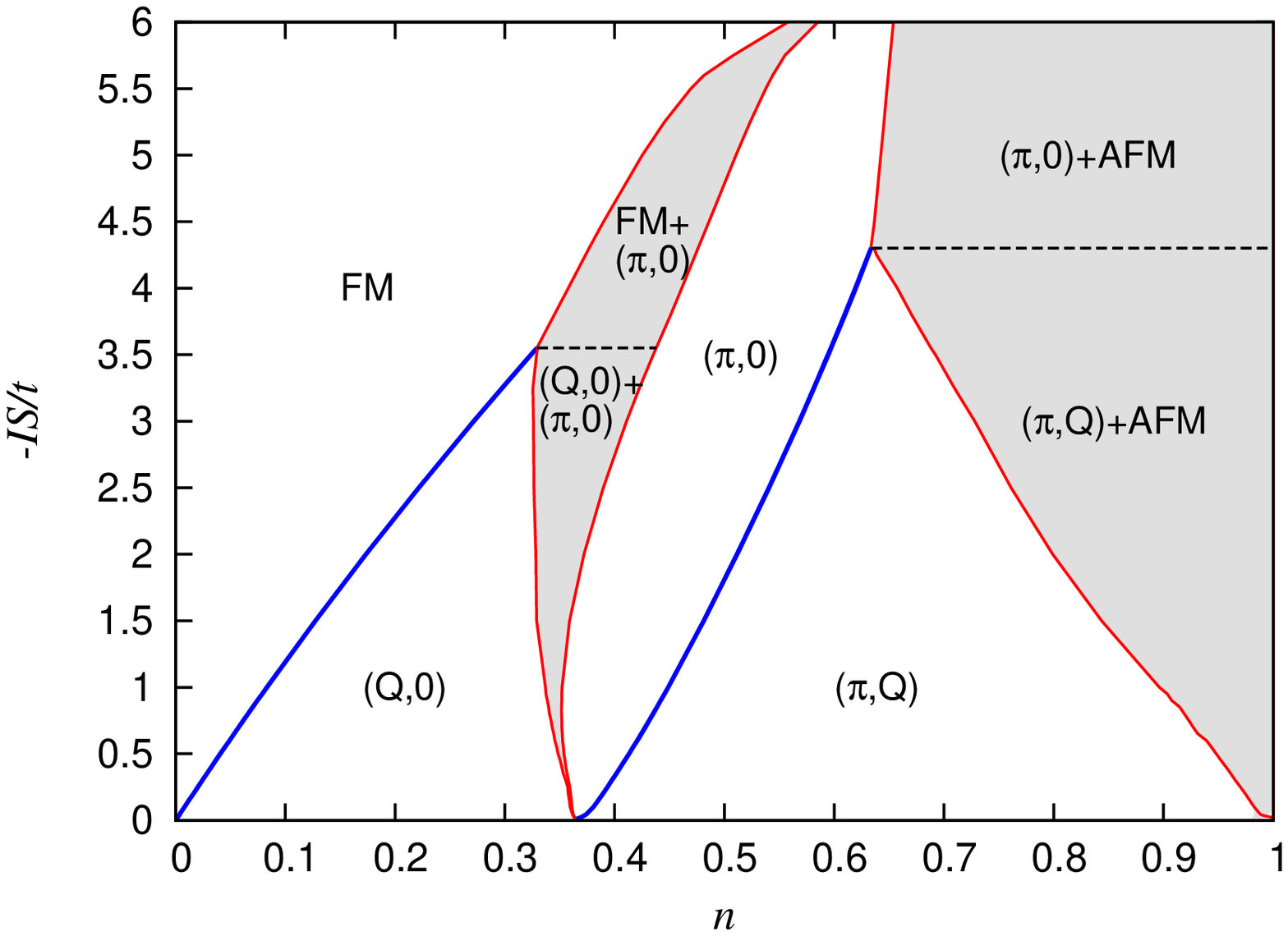}
\caption{
	(Color online)
	Ground state magnetic phase diagram of the Hubbard (upper panel, within (a) HFA~\cite{Igoshev10} and (b) SBA) and $s$-$d$ model (lower panel) for the square lattice 
	at $n<1$.
	The spiral phases are denoted according to the form of their wave vector.
	Filling shows the phase separation regions.
	Bold (blue) lines denote the second-order phase transitions.
	Solid (red) lines correspond to the boundaries between the regions of the homogeneous phase and phase
separation.
	Dashed (red) lines separate the PS regions corresponding to different phase pairs.
	The second-order phase transition boundary produced by the Stoner criterion 
	separating the ferromagnetic and paramagnetic regions, is shown by dash--dotted (violet) line,
	$\mathbf{Q}_{\rm AFM} = (\pi,\pi)$
}
\label{fig:sq}
\end{figure}

\begin{figure}[h!]
\includegraphics[width=0.49\textwidth]{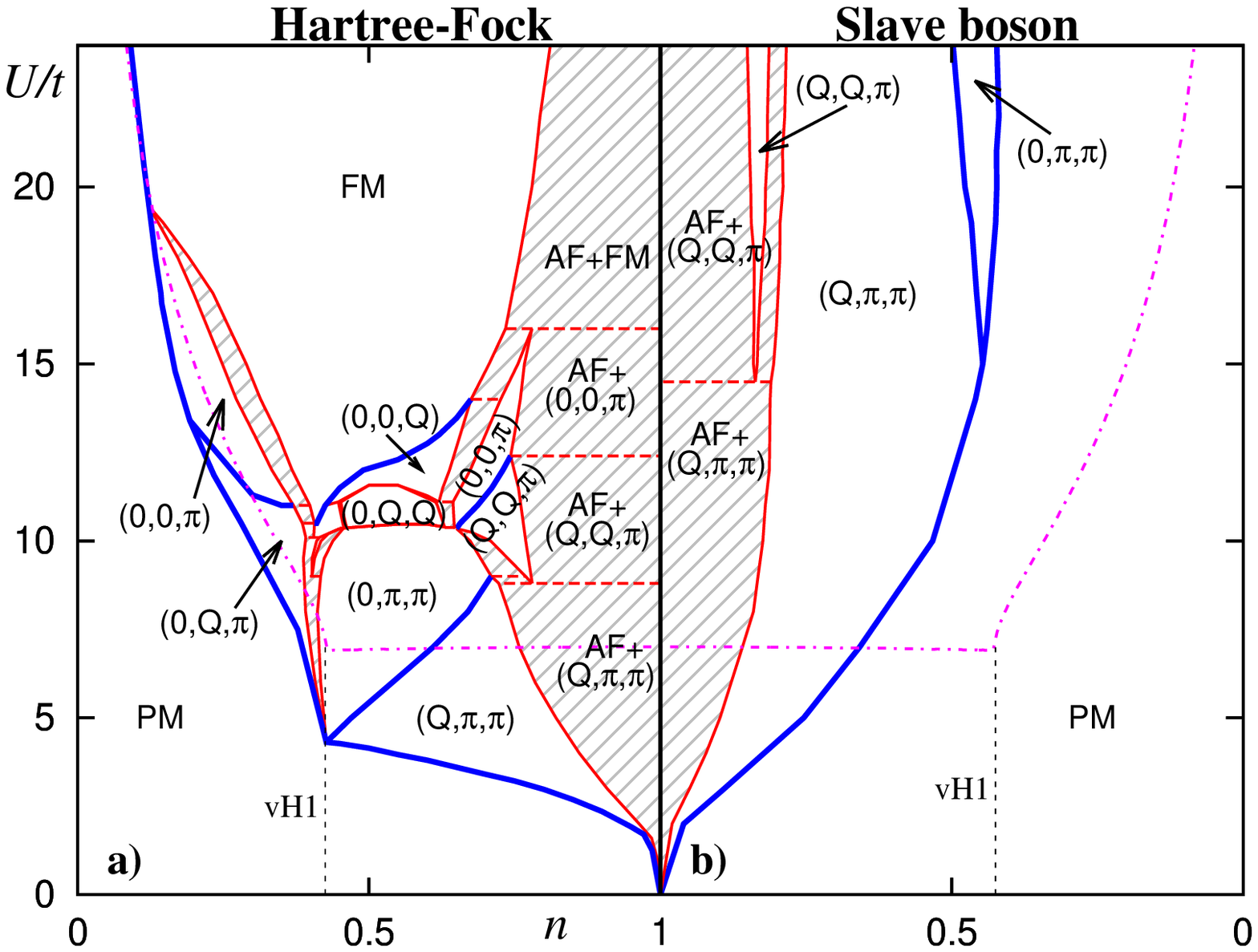}
\includegraphics[width=0.49\textwidth]{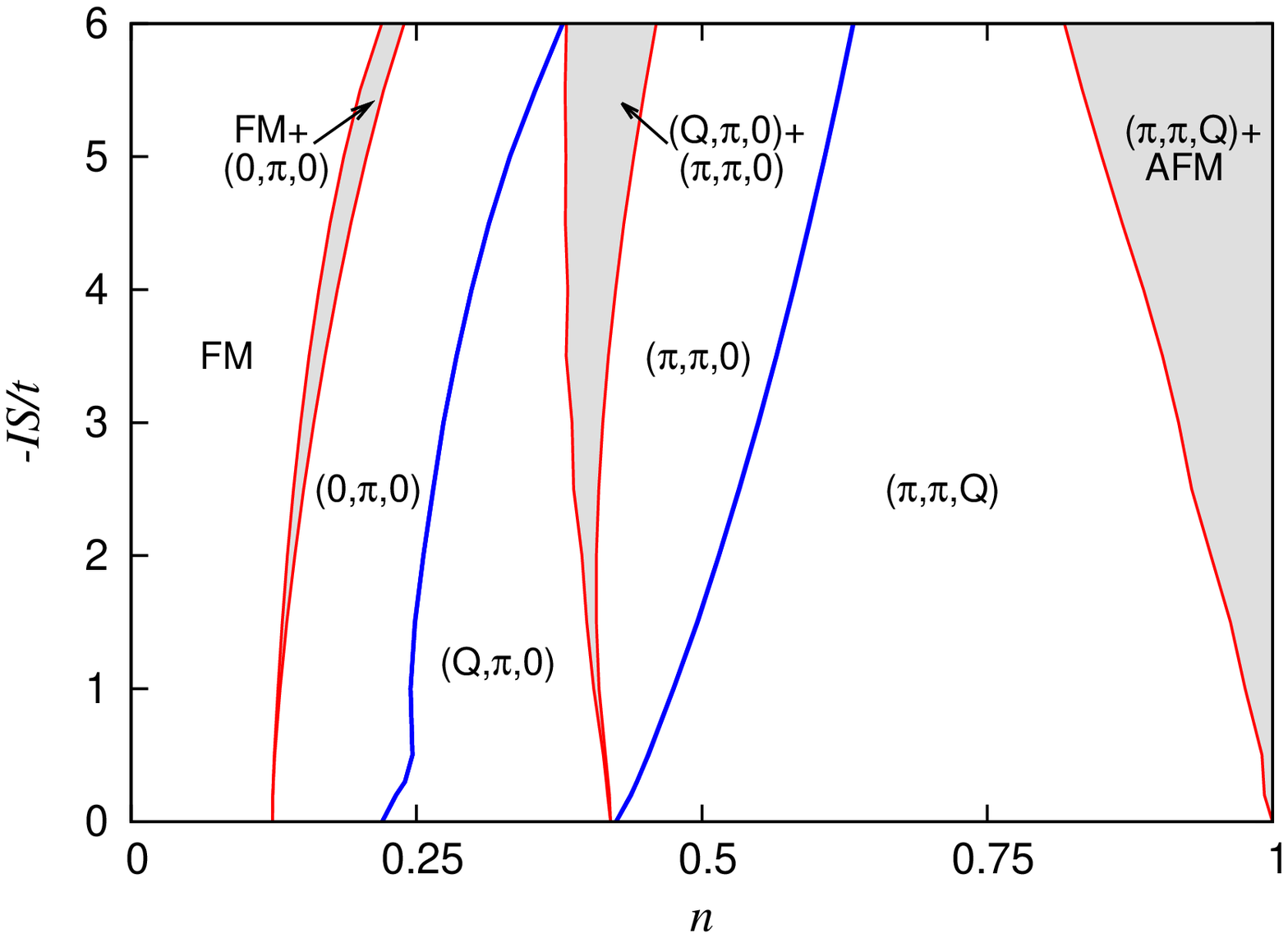}
\caption{
	(Color online)
	The phase diagrams for the simple cubic  lattice. Notations are the same as in Fig. \ref{fig:sq}, $\mathbf{Q}_{\rm AFM} = (\pi,\pi, \pi)$ 
}
\label{fig:sc}
\end{figure}

We present the results for the square~(Fig.~\ref{fig:sq}) and simple cubic~(Fig.~\ref{fig:sc}) lattices in the Hubbard within the HFA and SBA approximations and the $s$-$d$ exchange model.

One can see that  HFA yields the variety of spiral magnetic phases, as well as FM and AFM ones at sufficiently large $U$ generally and at any $U$ in the vicinity of half-filling. 
The account of correlations (SBA) 
leads to a noticeable suppression of  magnetically ordered states in comparison
with HFA: the corresponding concentration intervals in the phase diagram decrease strongly, and the variety of the spiral states disappears.
Besides that, in SBA there occurs a wide region of PM state which is a manifestation of correct treatment of the energy of doubles. 




For the Hubbard model on the square lattice  we have   
in the
vicinity of half-filling only the spiral $(Q,Q)$ (diagonal) phase, far away from half-filling
we have a FM phase for large $U/t$ and the parallel spiral phase ($(0,\pi)$ or $(Q,\pi)$) for moderate
$U/t$. Here $Q$ is a {\it number} depending on the model parameters.



The physical picture for the cubic lattice is  similar to that for the square lattice (see Fig.~\ref{fig:sc}a for a comparison of the results of HFA (a) and SBA (b)
approaches): the density value $n=1$ corresponding to the perfect AFM
nesting peculiarity (but not to a van Hove singularity) retains its crucial role.
The spiral magnetic phase with $\textbf{Q} = (Q,\pi)$ of the square lattice is
replaced by a spiral with ${\bf Q}=(Q,\pi,\pi)$.
The density of states 
has a van Hove (vH) 
singularity at the electron concentration $n_{\rm vH1}\approx0.425$. 
In contrast to the 2D case, the van Hove singularities are bounded and do not much affect the SBA diagram --- the correlation effects
strongly suppress magnetism at $n$ being far from half-filling, including the regions close to $n_{\rm vH}$'s.

It should be noted that the description of ferromagnetism at small $n$ requires the account of $T$-matrix renormalizations~\cite{Kanamori1963}. 

The ground state phase diagram for the $s$-$d$ model substantially differs from the results for the Hubbard model: the phases are strongly redistributed. 
The increasing of $|I|$ results in growing of ferromagnetic region. 
At small $|I|$ the wave vector of magnetic phase is specified by the position of the maximum of the Lindhardt function 
\begin{equation}\label{eq:sd_model_diagram}
	\chi^0_{\mathbf{Q}} = \frac1{N}\sum_{\kk}\frac{f_{\mathbf{k}+\mathbf{Q}/2}-f_{\mathbf{k}-\mathbf{Q}/2}}{t_{\mathbf{k}-\mathbf{Q}/2}-t_{\mathbf{k}+\mathbf{Q}/2}}
\end{equation}
calculated in the paramagnetic phase. 
An important difference of the square and sc lattice (2D and 3D) cases is the form of ferromagnetic region at small $n$. In the former case its width vanishes at $I\rightarrow0$, but in the latter case the width is finite and sufficiently large, the transition to the spiral $(0,\pi,0)$ phase being of the first order. 

For the square lattice the spiral phases are fully suppressed by FM and AFM regions at $|IS|\gtrsim6t$. 
For sc lattice the spiral phases turn out to be more stable. 
For small number of carriers in the AFM matrix ($1-n\ll 1$)  the phase separation between AFM and spiral ($(Q,\pi)$ for square lattice and $(Q,\pi, \pi)$ for sc lattice) phases is present at small interaction parameter. 

Within the mean-field approximation, the Hubbard model is equivalent to the $s$-$d$ model with the replacement $IS = Um/2$. However, the phase separation condition and description of the paramagnetic phase are different in these models.
Due to existence of localized moments,  ferromagnetic ordering is favorable 
already at small $|I|$, whereas within the Hubbard model it occurs at sufficiently large $U$ only.

Our calculations for the square lattice $s$-$d$ model are in a qualitative agreement with the Monte Carlo calculations for the classical $s$-$d$ model \cite{Hamada} and slave-fermion treatment of the Kondo lattice \cite{Lacroix} (phase separation and spirals are neglected in both the works). The spiral phases were taken into account in Ref. \cite{Costa}. 
However, the $(Q,Q)$ spiral phase obtained in this work turns out to be unstable with respect to phase separation into the N\'eel AFM and $(Q,\pi)$ spiral phases. 

With increasing $|I|$, the Kondo screening of the localized moments occurs, which is crucial for rare-earth compounds. 
This screening can be also included in our phase diagrams and will be considered elsewhere. 
The results for 2D and 3D cases are expected to be considerably different too.

\section{Acknowledgements}
\label{sec:akhnoledgments}
The research was carried out within the state assignment of FASO of Russia (theme ``Electron'' No. 01201463326). 
This work was supported in part by the Division of Physical Sciences and Ural Branch, Russian Academy of Sciences
(project no. 15-8-2-9, 15-8-2-12) and by the Russian Foundation for Basic Research (project no.
16-02-00995-a) and Act 211 Government of the Russian Federation 02.A03.21.0006. 
The main amount of calculations was performed using the ``Uran'' cluster of IMM UB RAS.





\end{document}